\documentclass[12pt]{article}
\usepackage{amsmath,amssymb}
\usepackage{graphicx}
\usepackage{cite,fancyhdr}
\pdfoutput=1

\usepackage[affil-it]{authblk}
\usepackage[left=2cm,right=2cm,top=2cm,bottom=2cm,a4paper]{geometry}

\allowdisplaybreaks

\begin{document}

\setcounter{footnote}{0}
\setcounter{figure}{0}
\setcounter{table}{0}

\title{\bf \Large 
Implications of the 750 GeV Diphoton Excess in Gaugino Mediation}
\author[1]{{\normalsize Chengcheng Han}}
\author[1]{{\normalsize Tsutomu T. Yanagida}}
\author[2]{{\normalsize Norimi Yokozaki}}

\affil[1]{\small Kavli Institute for the Physics and Mathematics of the Universe (WPI),   \authorcr
  {\it Todai Institutes for Advanced Study, University of Tokyo,
  Kashiwa 277-8583, Japan}}
\affil[2]{\small Istituto Nazionale di Fisica Nucleare, Sezione di Roma, \authorcr {\it Piazzale Aldo Moro 2, I-00185 Rome, Italy}
}

\date{}

\maketitle

\thispagestyle{fancy}
\rhead{IPMU16-0016}
\cfoot{\thepage}
\renewcommand{\headrulewidth}{0pt}

\begin{abstract}
\noindent

The 750 GeV diphoton excess reported by ATLAS and CMS indicates the presence of several pairs of the vector-like matter multiplets around TeV scale. If that is the case, radiative corrections from the $SU(3)$ gauge interaction significantly change from those of MSSM, 
and the infrared-free nature of the gauge interaction leads to characteristic SUSY mass spectra: a ratio of a  squark mass to the gluino mass, and scalar trilinear couplings are enhanced at the low-energy scale. Consequently, even in gaugino mediation models, the Higgs boson mass of 125 GeV is explained with the fairly light gluino of 2-3 TeV, which can be accessible at the LHC. 
\end{abstract}

\clearpage

\section{Introduction}
Gaugino mediation~\cite{gm1, gm2} provides an attractive framework of mediating supersymmetry (SUSY) breaking effects to the observable sector, since this framework is free from the SUSY flavor changing neutral current problem. 
In gaugino mediation, only gaugino masses and a $\mu$-parameter (and the Higgs $B$-term) are assumed to be non-vanishing at the high energy scale 
and all SUSY particle masses at the low-energy scale are determined mostly by the gaugino masses at the high energy scale. Therefore, this framework is very predictive.
Provided that particle contents below the grand unified theory (GUT) scale are those in minimal supersymmetric standard model (MSSM), the observed Higgs boson mass of 125 GeV is explained with the gluino mass larger than 5-6 TeV~\cite{Moroi:2012kg}, which is unfortunately beyond the reach of the LHC experiments.

However, the diphoton excess recently reported by ATLAS~\cite{diphoton_atlas} and CMS collaborations~\cite{diphoton_cms} indicates the presence of several pairs of the vector-like matter multiplets around or below the TeV scale, in addition to MSSM matter contents~\cite{vector-diphoton_refs, franceschini}. To explain the diphoton excess 
with the cross section of $\simeq$ 5\,fb,  rather large number of vector-like pairs is required as long as 
relevant Yukawa couplings are not larger than unity. 
It has been shown that the four pairs of leptonic matters with masses of $\simeq$ 400 GeV 
and colored ones with masses of $\simeq$ 800 GeV can reproduce the observed diphoton signal 
under the conditions that those vector-like matters form complete $SU(5)$ multiplets and the perturbativity of the relevant couplings is maintained up to the GUT scale~\cite{vector-diphoton_refs}. 

In fact, vector-like multiplets around TeV scale significantly change the SUSY mass spectrum at the low-energy,
due to remarkable changes of gauge and gaugino beta-functions~\cite{Moroi:2012kg}: at the low-energy, ratios of sfermion masses to gaugino masses become much larger than those evaluated in MSSM. Moreover, the stop trilinear coupling becomes large. As a result, the Higgs boson mass of $\simeq$ 125 GeV is explained with the gluino mass accessible at the LHC experiment.

In this paper, we revisit our previous study in the light of the diphoton excess, and show that the gluino mass is likely to be lighter than 2-3 TeV. We take account of threshold corrections to gauge couplings from the vector-like matter multiplets  lighter than 1 TeV, and evaluate MSSM mass spectra using two-loop renormalization group equations (RGEs), which are required due to large couplings at the GUT scale.

\section{SUSY explanation of the diphoton excess}
One 	of the plausible models to explain the observed diphoton excess is a model 
which contains several pairs of vector-like fermions. 
The vector-like fermions couple to a singlet scalar boson $\mathcal{S}$ of 750 GeV. 
As a SUSY realization, we consider the following superpotential:
\begin{eqnarray}
W =  W_{\rm MSSM} + 
\lambda_D S D' \bar D' + \lambda_L S L' \bar L'  + M_D D' \bar D' + M_L L' \bar L' + \frac{M_S}{2} S^2,
\end{eqnarray}
with soft SUSY breaking terms
\begin{eqnarray}
-\mathcal{L}_{\rm soft} &=& -\mathcal{L}_{\rm soft}^{\rm MSSM} + 
m_{S}^2 |S|^2 + m_{D'} |D'|^2 + m_{\bar D'} |\bar D'|^2 
+ m_{L'} |D'|^2 + m_{\bar L'} |\bar L'|^2 \nonumber \\
&+& \left( A_{L'} \lambda_L S L' \bar L'  + A_{D'} \lambda_D S D' \bar D'  + h.c. \right) \nonumber \\
&+& \left( B_D M_D D' \bar D' +  B_L M_L L' \bar L'  
+  B_S \frac{M_S}{2} S^2 + h.c. \right),
\end{eqnarray}
where $W_{\rm MSSM}$ and $\mathcal{L}_{\rm soft}^{\rm MSSM}$ are the superpotential and soft SUSY breaking terms of the MSSM, respectively, $S$ is a gauge singlet chiral superfield, $\bar D'$ and $D'$ have $SU(3)_c \times SU(2)_L \times U(1)_Y$ charges of $({\bf \bar 3}, {\bf 1}, 1/3)$ and $({\bf 3},{\bf 1},-1/3)$, and $L'$ and $\bar L'$ have $({\bf 1}, {\bf 2}, -1/2)$ and $({\bf 1}, {\bf 2}, 1/2)$. The vector-like multiplets belong to complete $SU(5)$ multiplets as ${\bf \bar 5}= (L', \bar D')$  and ${\bf 5} = (\bar L', D')$. We introduce four pairs of ${\bf 5}$ and ${\bf {\bar 5}}$. Here, $M_D$, $M_L$ and $\lambda_D$ are taken to be real positive without a loss of generality.

One of bosons, $\mathcal{S}$, in the chiral superfield $S$ is dominantly produced by the gluon fusion process 
%
%
and it subsequently decays to diphoton radiatively. 
Using narrow width approximation, the cross section of $pp \to \mathcal{S} \to \gamma \gamma$ is estimated as
\begin{eqnarray}
\sigma(pp \to \mathcal{S} \to \gamma\gamma) &\simeq& K \cdot \, \frac{\pi^2}{8 m_{\mathcal{S}}} \frac{1}{s} \, 
\Gamma( \mathcal{S} \to gg) {\rm Br}( \mathcal{S} \to \gamma\gamma) C_{gg}, \nonumber \\
C_{gg} &=& \int_0^1 d x_1 \int_0^1 d x_2 f_g(x_1) f_g (x_2) \delta(x_1 x_2 - m_{\mathcal{S}}^2/s),
\end{eqnarray}
where $K$ is a $K$ factor, $\sqrt{s}=13$\,TeV and  $m_{\mathcal{S}}=750$\,GeV.
Using {\tt MSTW2008NNLO}~\cite{mstw2008} with the factorization scale of $0.5 m_{\mathcal{S}}\, (m_{\mathcal{S}})$, $C_{gg} \approx 1904\, (1736)$.
Then,
\begin{eqnarray}
\sigma(pp \to \mathcal{S} \to \gamma\gamma) \approx K \cdot 7.2\, (6.6)\, {\rm fb} \, 
\left( \frac{\Gamma(\mathcal{S} \to \gamma\gamma)}{10^{-3}\, {\rm GeV}} \right),
\end{eqnarray}
where we have used ${\rm Br}( \mathcal{S} \to \gamma\gamma) \approx \Gamma( \mathcal{S} \to \gamma\gamma)/\Gamma( \mathcal{S} \to gg)$. 

In our setup, the radiatively generated $B_S$ is large and positive ($B_S \sim 5 {\rm TeV}$), leading to the large mass splitting of two real states in $S$. Therefore, the only lighter state contributes the diphoton cross section.
Provided $M_S$ is real positive, the lighter state is CP-odd,\footnote{
The decay width of the CP-odd state is larger than that of the CP-even state.
See e.g. \cite{Djouadi:2005gj} for the difference of the form factors for the CP-even and odd states.
} and the cross section is
\begin{eqnarray}
\sigma(pp \to \mathcal{S} \to \gamma\gamma) \approx 4.1 \, (3.8) \, {\rm fb} \, \left( \frac{K}{1.5} \right),
\end{eqnarray}
where we take $M_L=400$ GeV, $M_D=800$ GeV, $\lambda_L= \lambda_D=0.4$.\footnote{
We have found that three pairs of ${\bf 5}+\bar{\bf 5}$ with a similar mass spectrum and couplings can marginally explain  the diphoton excess without the divergences of the coupling constants below the GUT scale.
}
 The typical value of the K factor is around 1.5~\cite{franceschini}. With these values, $\lambda_D$ and $\lambda_L$ do not unify at the GUT scale. However, this may not be a problem since such a disparity may be easily generated by GUT symmetry breaking terms.

\begin{figure}[!t]
\begin{center}
\includegraphics[scale=1.1]{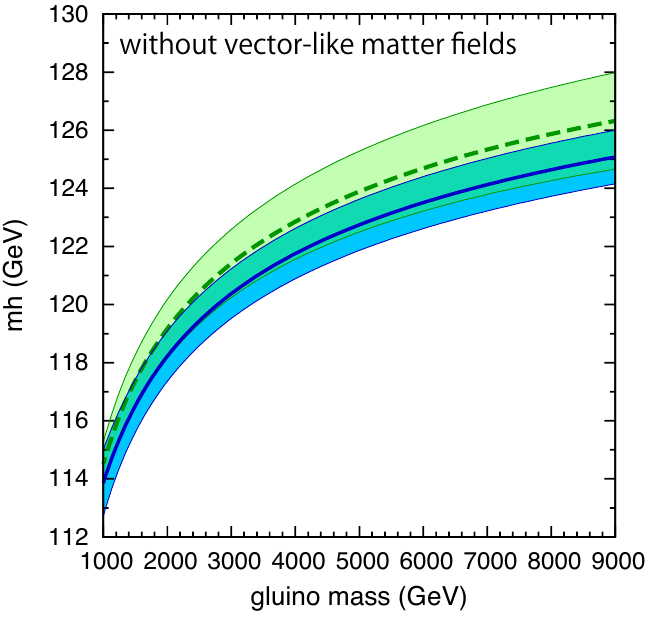}
\caption{The Higgs boson mass as a function of the gluino mass without the extra matters. 
Here, $\alpha_s(m_Z)=0.1185$, $m_t({\rm pole})=173.34$ GeV and $\tan\beta=25$.
}
\label{fig:mh_mssm}
\end{center}
\end{figure}

\section{SUSY mass spectra in gaugino mediation}

In this section, we evaluate SUSY mass spectra with the vector-like matter fields lighter than one TeV in gaugino mediation. In gaugino mediation models, we assume the K{\" a}hler potential is the sequestered form~\cite{gm1, gm2, sequestered} or the no-scale form~\cite{noscale}, 
and a SUSY breaking field only couples to gauge field-strength superfields directly. Then, at the high energy scale, sfermion masses vanish at the leading order and only gaugino masses and $\mu$-term are not suppressed. The Higgs $B$-term, $B_H$, may or may not be suppressed, depending on a setup.\footnote{
The Higgs $B$-term vanishes in e.g. Scherk-Schwarz SUSY breaking~\cite{sssb} and the gravitational SUSY breaking~\cite{gravsb}. In those cases, the $F$-term of the chiral compensator field vanishes at the leading order.
} Gaugino mediation models are parametrized 
with $M_{1/2}$, $\mu(M_{\rm GUT})$ and $B_H(M_{\rm GUT})$ or more conveniently,
\begin{eqnarray}
M_{1/2}, \, \tan\beta, \, {\rm sign}(\mu),
\end{eqnarray}
taking into account conditions for the correct electroweak symmetry breaking (EWSB). We identify the cut-off scale with the GUT scale, $M_{\rm GUT}$, and $M_{1/2}$ is the universal gaugino mass at $M_{\rm GUT}$. If the Higgs $B$-term, $B_H$, vanishes at the high energy scale, $\tan\beta$ and ${\rm sign}(\mu)$ are not free-parameters  but predictions. This possibility will be revisited later. 

Before discussing the impact of the light vector-like matters, 
let us briefly show the expected gluino mass in gaugino mediation without vector-like matters. 
In Fig.~\ref{fig:mh_mssm}, we plot the mass of the SM-like Higgs boson as a functions of the physical gluino mass.
SUSY mass spectra are computed using {\tt Suspect v2.43} package~\cite{suspect}. 
The blue solid (green dashed) line shows the computed Higgs boson mass using {\tt SusyHd v1.02} code~\cite{susyhd} ({\tt FeynHiggs v.2.11.3} code~\cite{feynhiggs}). We take $m_t({\rm pole})=173.34$ GeV, $\alpha_s(m_Z)=0.1185$, $\tan\beta=25$ and $\mu>0$.
The blue and green bands show the uncertainties of the theoretical calculations, including the error of the measured top mass, $\pm 0.76$ GeV~\cite{topmass}. Although we see the difference between the results of the two different codes, 
the Higgs boson mass of 125 GeV requires a rather heavy gluino of 5-8 TeV without extra vector-like multiplets. 

\vspace{12pt}
However, as shown in Ref.~\cite{Moroi:2012kg}, 
the existence of the light vector-like matter fields have significant effects on SUSY particle masses in gaugino mediation: 1) squark/gluino mass ratios as well as 2) $A$-term/squark mass ratio are enhanced. 
As a result, the predicted gluino mass becomes significantly smaller than that without vector-like matter fields 
since the observed Higgs boson mass does not require the heavy gluino.

At the one-loop level, the change of the beta-functions is simply given by
\begin{eqnarray}
\beta_i  &=& (\beta_i)_{\rm MSSM} + \frac{N_5}{16\pi^2} g_i^3 \, , \nonumber \\
\beta_{M_i}  &=& (\beta_{M_i})_{\rm MSSM} + \frac{2 N_5}{16\pi^2} M_i g_i^2 \,,
\end{eqnarray}
where $N_5(=4)$ is the number of the vector-like matter multiplets, 
$M_1$, $M_2$ and $M_3$ are the bino, wino and gluino mass, 
$g_1$, $g_2$ and $g_3$ are gauge couplings of $U(1)_Y$, $SU(2)_L$ and $SU(3)_c$,
and $\beta_i$ and $\beta_{M_i}$ are beta-functions for gauge couplings and gaugino masses, respectively. 
Since the SUSY breaking masses of $L'$, $\bar L'$, $D'$ and $\bar D'$ are much larger than $M_L$ and $M_D$, 
the radiative corrections to the gauge couplings from the scalar components and fermion components should be taken into account separately. 
The threshold corrections from the vector-like matter multiplets can be included as~\footnote{
Threshold corrections from SUSY particles are included in the numerical calculations, which are important as well.
} 
\begin{eqnarray}
g_1^{-2}(m_{\rm SUSY}) &\to& g_1^{-2}(m_{\rm SUSY}) - \frac{N_5}{8\pi^2} \frac{4}{5} 
\left[ \frac{1}{2} \ln \frac{m_{\rm SUSY}}{M_L} + \frac{1}{3} \ln \frac{m_{\rm SUSY}}{M_D} \right] \nonumber \\
&-&  \frac{N_5}{8\pi^2} \frac{2}{5} \left[ \frac{1}{4} \ln \frac{m_{\rm SUSY}}{m_{L'_1}} + \frac{1}{6} \ln \frac{m_{\rm SUSY}}{m_{D'_1}} + \frac{1}{4} \ln \frac{m_{\rm SUSY}}{m_{L'_2}} + \frac{1}{6} \ln \frac{m_{\rm SUSY}}{m_{D'_2}} \right]  \, ,\nonumber \\
g_2^{-2}(m_{\rm SUSY}) &\to& g_2^{-2}(m_{\rm SUSY}) - \frac{N_5}{8\pi^2}  
\left[ \frac{2}{3} \ln \frac{m_{\rm SUSY}}{M_L} + \frac{1}{6} \ln \frac{m_{\rm SUSY}}{m_{L'_1}} 
+ \frac{1}{6} \ln \frac{m_{\rm SUSY}}{m_{L'_2}} \right] \, ,\nonumber \\
g_3^{-2}(m_{\rm SUSY}) &\to& g_3^{-2}(m_{\rm SUSY}) - \frac{N_5}{8\pi^2} 
\left[ \frac{2}{3} \ln \frac{m_{\rm SUSY}}{M_D} + \frac{1}{6} \ln \frac{m_{\rm SUSY}}{m_{D'_1}}
+ \frac{1}{6} \ln \frac{m_{\rm SUSY}}{m_{D'_2}} \right] \, , \label{eq:thc}
\end{eqnarray}
where $m_{\rm SUSY}$ is a mass scale of the SUSY particles, and $m_{L'_i}$ and $m_{D'_i}$ are mass eigenvalues of the scalar components taking into account mass splittings from $B_L$ and $B_D$.

Now, we discuss the enhancements of sfermion mass and $A$-terms. Although two-loop RGEs are important to evaluate SUSY mass spectra as shown below, the enhancements of $A$-term and sfermion masses can be understood qualitatively by solving one-loop RGEs.
The $A$-terms of the first and second generation up-type squarks are 
\begin{eqnarray}
A_u (m_{\rm SUSY}) \simeq -\sum_i \frac{c_i}{2} \left( \frac{g_i^4(m_{\rm SUSY})}{8\pi^2} \ln \frac{M_{\rm GUT}}{m_{\rm SUSY}} \right)
 \left(1 - \frac{b_i}{8\pi^2} g_i^2(m_{SUSY}) \ln \frac{M_{\rm GUT}}{m_{\rm SUSY}}\right)^{-1} \tilde M,
\end{eqnarray}
where $(c_1, c_2, c_3)=(26/15, 6, 32/3)$,  $(b_1, b_2, b_3)=(33/5 + N_5, 1 + N_5, -3 + N_5)$ and $\tilde M = M_3/g_3^2 \simeq M_2/g_2^2 \simeq M_1/g_1^2$. 
Neglecting $U(1)_Y$ and $SU(2)_L$ contributions and threshold corrections in Eq.(\ref{eq:thc}), 
we have an enhancement of 
\begin{eqnarray}
\frac{A_u (m_{\rm SUSY})}{(A_u(m_{\rm SUSY}))_{\rm MSSM}} \simeq \frac{1  + \frac{3}{8\pi^2} g_3^2(m_{SUSY}) \ln \frac{M_{\rm GUT}}{m_{\rm SUSY}}}{1 - \frac{1}{8\pi^2} g_3^2(m_{SUSY}) \ln \frac{M_{\rm GUT}}{m_{\rm SUSY}}} \sim 3.4\, ,
\end{eqnarray}
for the fixed gluino mass at $m_{\rm SUSY}$ where $m_{\rm SUSY} \simeq 3.5$ TeV and $N_5=4$.
Similarly, the squark mass becomes larger than that of MSSM as 
\begin{eqnarray}
\frac{m_{Q}^2(m_{\rm SUSY})}{(m_Q^2(m_{\rm SUSY}))_{\rm MSSM}}  &\simeq&
\frac{(1 - \frac{1}{8\pi^2} g_3^2(m_{SUSY}) \ln \frac{M_{\rm GUT}}{m_{\rm SUSY}})^{-2}+(1 - \frac{1}{8\pi^2} g_3^2(m_{SUSY}) \ln \frac{M_{\rm GUT}}{m_{\rm SUSY}})^{-1}}{(1  + \frac{3}{8\pi^2} g_3^2(m_{SUSY}) \ln \frac{M_{\rm GUT}}{m_{\rm SUSY}})^{-2}+(1  + \frac{3}{8\pi^2} g_3^2(m_{SUSY}) \ln \frac{M_{\rm GUT}}{m_{\rm SUSY}})^{-1}} \nonumber \\
&\sim& 5.9 \,.
\end{eqnarray}
Accordingly, 
\begin{eqnarray}
\frac{A_u^2(m_{\rm SUSY})/m_Q^2(m_{\rm SUSY})}{ \left( A_u^2(m_{\rm SUSY})/m_Q^2(m_{\rm SUSY}) \right)_{\rm MSSM}} \sim 1.9,
\end{eqnarray}
which is crucial for the enhancement of the Higgs boson mass.

\begin{figure}[!t]
\begin{center}
\includegraphics[scale=0.9]{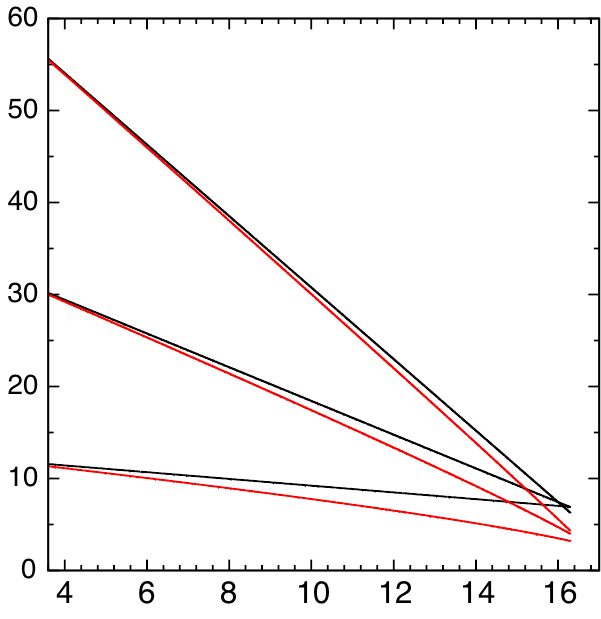}
\hspace{2pt}
\includegraphics[scale=0.9]{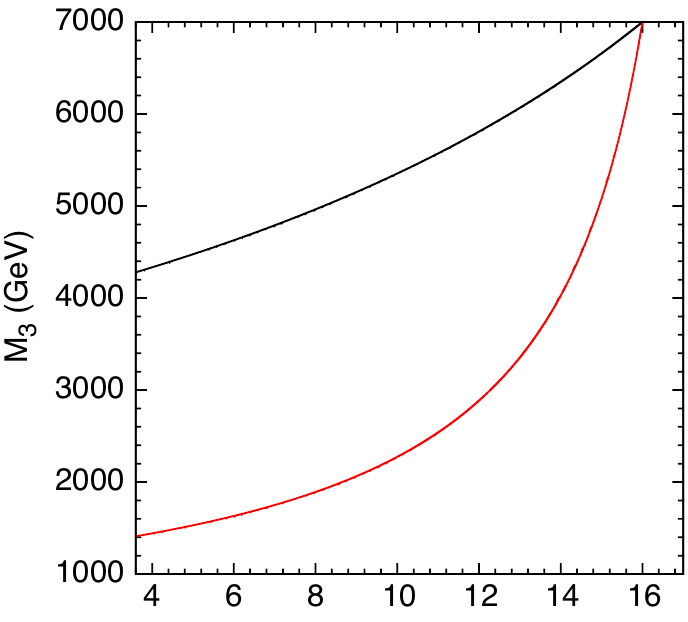}
\caption{The RGE runnings of the gauge couplings ($\alpha_{1,2,3}^{-1}$) and gluino mass. 
The horizontal axis is $\log_{10}(Q_R/{\rm GeV})$, where $Q_R$ is a renormalization scale.
The red (black) lines show the results at the two-loop (one-loop) level. Here, 
$M_{1/2}=7$\,TeV, $\tan\beta=10$, 
$M_L=400$\,GeV, $M_D=800$\,GeV, 
$\lambda_D=\lambda_L=0.4$, 
$\alpha_s(M_Z)=0.1185$ and $m_t({\rm pole})=173.34$ GeV.
}
\label{fig:gauge}
\end{center}
\end{figure}

\begin{figure}[!t]
\begin{center}
\includegraphics[scale=1.0]{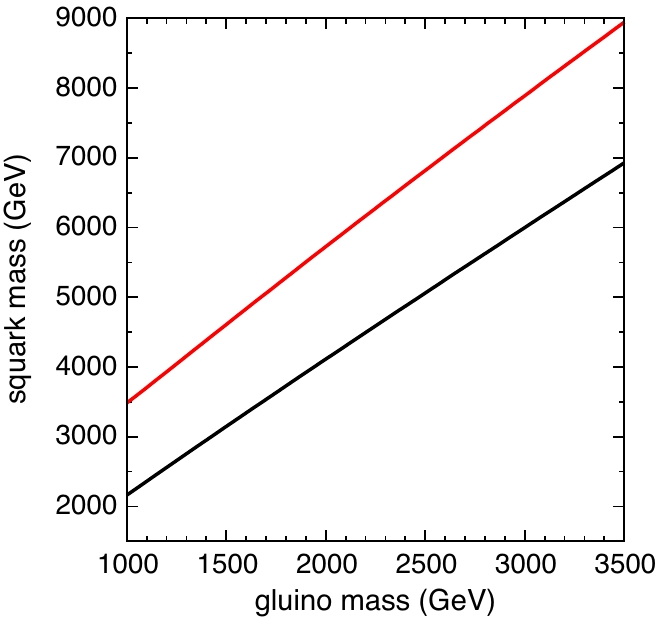}
\caption{The squark mass as a function of the gluino mass. The red (black) line shows the result at the two-loop (one-loop) level. The model parameters except for $M_{1/2}$ are the same as in Fig.~\ref{fig:gauge}.
}
\label{fig:msq}
\end{center}
\end{figure}

Although the above discussion is almost correct, 
two-loop RGEs are required to be included since the gauge couplings are quite large at the high energy scale. A set of the two-loop RGEs is shown in Appendix \ref{ap:rges}. 
In Fig.\,\ref{fig:gauge}, 
the RGE runnings of the gauge couplings and gluino mass with one- and two-loop RGEs are shown as functions of the renormalization scale.
The model parameters are taken as $M_L=400$\,GeV, $M_D=800$\,GeV,  $\tan\beta=10$, $\alpha_s(M_Z)=0.1185$,
and $m_t({\rm pole})=173.34$ GeV, and $\lambda_D$ and $\lambda_L$, are fixed to be 0.4 at the weak scale. The universal gaugino mass at the GUT scale ($10^{16}$\,GeV) is taken as $M_{1/2} = 7$\,TeV. In the plot, squark masses are around 5 TeV and gluino mass is around 2 TeV. 
The deviation of the gluino masses at the one and two-loop runnings is quite large. The large correction comes from the following terms:
\begin{eqnarray}
\beta_{g_3} \ni \frac{g_3^5}{(16\pi^2)^2} \frac{178}{3} \, , \ \,
\beta_{M_3} \ni \frac{g_3^4}{(16\pi^2)^2} \frac{712}{3} M_3 \, . 
\end{eqnarray}
In fact, without the above contributions, the RGE running at the two-loop level is not much different from that evaluated at the one-loop level. 

\begin{figure}[!t]
\begin{center}
\includegraphics[scale=1.1]{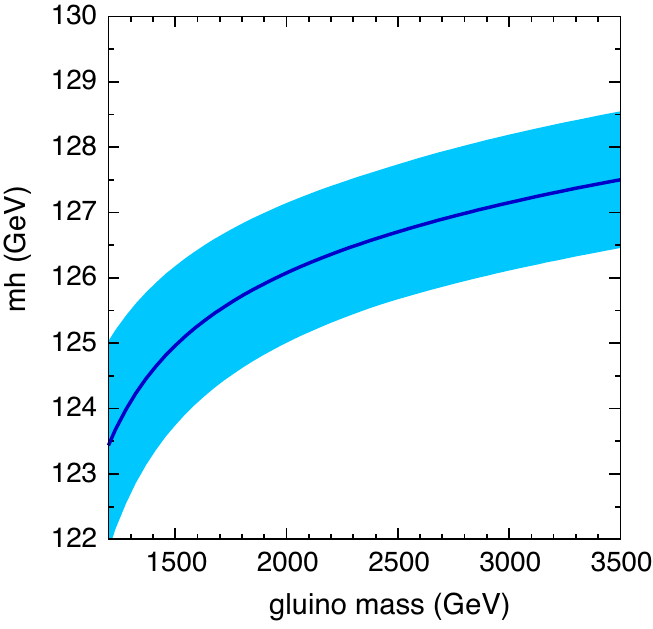}
\caption{The Higgs boson mass as a function of the gluino mass. 
The blue band shows the theoretical uncertainty of the Higgs boson mass calculation.
Here, $\tan\beta=9$ and other parameters are the same as in Fig.~\ref{fig:gauge}. 
}
\label{fig:mh}
\end{center}
\end{figure}

The difference of the RGE evolutions at the one- and two-loop level affects the squark masses at the low-energy scale. In Fig.~\ref{fig:msq}, we show the squark mass 
(the mass of the $SU(2)_L$ singlet up-type squark) as a function of the physical gluino mass using one- and two-loop RGEs. 
The squark mass evaluated at the two-loop level is larger by about 1 TeV than the one-loop computation, which obviously affects the Higgs boson mass calculation. 
The calculated Higgs boson mass is shown in Fig.~\ref{fig:mh} for $\tan\beta=9$. We compute the Higgs boson mass using {\tt SusyHd}, with the theoretical uncertainty including the error of the measured top mass. The predicted gluino mass is in a range of 1.2-2.2\, TeV, including the theoretical uncertainty, which is accessible at the LHC.

\begin{figure}[!t]
\begin{center}
\includegraphics[scale=1.3]{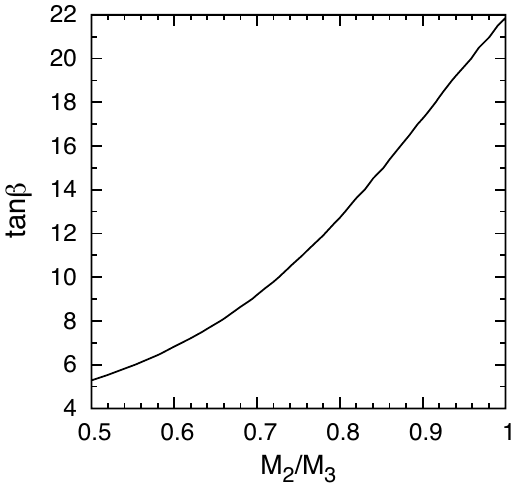}
\caption{Predictions of $\tan\beta$ for $M_1=M_3=7000$ GeV with $B_{H}(M_{\rm GUT})=0$.
The other parameters are the same as in Fig.~\ref{fig:gauge}. 
}
\label{fig:tanb}
\end{center}
\end{figure}

As mentioned earlier, the Higgs $B$-term vanishes at the high energy scale in some setups, 
and $\tan\beta$ is not a free parameter in those cases. If the gaugino masses are taken to be universal at the GUT scale, $\tan\beta$ is predicted to be 20-22, slightly depending on the SUSY scale. However, non-universal gaugino masses arise in e.g. models of product group unification~\cite{pgu}, where the doublet-triplet splitting problem is naturally solved. With non-universal gaugino masses, the prediction of $\tan\beta$ is in a wider range. 
The predicted $\tan\beta$ in the case of non-universal gaugino masses are shown in Fig.~\ref{fig:tanb}. 
We have taken $M_1=M_3$ at the GUT scale but $M_2$ to be free.

\begin{table*}[!t]
\caption{\small Mass spectra in sample points. Here, $\lambda_D=0.45$. 
The other parameters not shown in the table are the same as in Fig.~\ref{fig:gauge}.
}
\label{tab:sample}
\begin{center}
\begin{tabular}{|c||c|c|c|c|}
\hline
Parameters & Point {\bf I} & Point {\bf II}  & Point {\bf III}\\
\hline
$M_{3}$ (GeV) & 7500  & 8500  & 7800 \\
$M_{1}/M_3$  & 1  & 1  & 0.793\\
$M_{2}/M_3$  & 1  & 1  &  0.6\\
$M_{\rm GUT}$ (GeV)& $10^{16}$  & $10^{16}$  & $10^{16}$\\
$\lambda_L$&  0.45  & 0.45 & 0.45\\
\hline
$\tan\beta$ & 8 & 6  & 6.7\\
$\mu$ (GeV) & 5570 & 6470  & 5930\\
$A_t$ (GeV) & -8320 & -9510 & -8080\\
\hline
%
Particles & Mass (GeV) & Mass (GeV)& Mass (GeV) \\
\hline
$\tilde{g}$ & 2050 & 2490 &2210  \\
$\tilde{q}$ & 5790-6190 & 6740-7210 & 6120-6130\\
$\tilde{t}_{2,1}$ & 5040, 2880 & 5920, 3490 & 5010, 3630\\
$\tilde{\chi}_{2,1}^\pm$ & 5550, 1060 & 6450, 1260  & 5910, 571\\
$\tilde{\chi}_4^0$ & 5550 & 6450 & 5910\\
$\tilde{\chi}_3^0$ & 5550 & 6450 & 5910 \\
$\tilde{\chi}_2^0$ & 1060 & 1260 & 571 \\
$\tilde{\chi}_1^0$ & 684 & 797  & 543\\
$\tilde{e}_{L, R}(\tilde{\mu}_{L, R})$ & 3480, 2170 & 2880, 1650  & 1750, 1530\\
$\tilde{\tau}_{2,1}$ & 3020, 1880& 3470, 2150 & 1750, 1520\\
$H^\pm$ & 6350 & 6410 & 6230\\
$h_{\rm SM\mathchar`-like}$ & 125.8 &  125.0  & 125.2 \\
\hline
\end{tabular}
\end{center}
\end{table*}

Finally, we present some mass spectra in our gaugino mediation model in Table \ref{tab:sample}. 
The Higgs boson mass is computed using {\tt SusyHd}, and the SUSY mass spectra are computed using {\tt Suspect} package with modifications of RGEs 
and an inclusion of Eq.(\ref{eq:thc}).
In the points {\bf I} and {\bf II}, the gaugino masses at the GUT scale are taken to be universal. 
In the point {\bf III}, the gaugino masses at the GUT scale are non-universal, and $B_H(M_{\rm GUT})=0$ is imposed 
and $\tan\beta$ shown in the Table is determined by the conditions for the correct electroweak symmetry breaking. 
In the point {\bf III} the relic abundance of the lightest neutralino is consistent with the observed value, $\Omega_{\rm LSP} h^2 \simeq 0.12$, thanks to bino-wino coannihilation. Here, the relic abundance is calculated using {\tt micrOMEGAs}~\cite{micromegas}.\footnote{
Non-perturbative effects change $\Omega_{\rm LSP} h^2$ slightly~\cite{coannihilation2}.
}

\section{Conclusion and discussion}
We have studied impacts of the light vector-like matter multiplets on the sparticle masses especially on the gluino mass in gaugino mediation, in the light of the diphoton excess. The existence of the light vector-like matter fields enhances trilinear couplings and squark/gluino mass ratios at the infrared scale. Consequently, the observed Higgs boson mass of $\simeq$ 125 GeV is explained with a light gluino: the gluino mass is likely to be smaller than 2-3 TeV unless $\tan\beta$ is small as $\mathcal{O}(1)$. 
The predicted gluino is fairly light and the LHC may soon discover the gluino in a final state with multi-jets 
and missing transverse momentum.

In cases of the universal gaugino masses, the abundance of the bino-like neutralino is too large compared to the observed dark matter abundance. Therefore, the small $R$-parity violation may be required to avoid the over-closure of the universe.
However, the non-universal gaugino masses at the GUT scale are expected to arise in models of the product group unification, where the doublet-triplet splitting problem is naturally solved. 
If the wino mass at the high energy scale is smaller than the bino mass (and the gluino mass) 
and the bino and wino masses are mildly degenerate at the weak scale, 
the relic abundance of the lightest neutralino can be consistent with the observed dark matter abundance, thanks to the bino-wino coannihilation.  The correct electroweak symmetry breaking is explained with the vanishing Higgs $B$-term at the high energy scale in this case.

In our paper, we have introduced SUSY invariant mass parameters for the singlet and vector-like matter fields. 
However, these mass parameters are not necessarily required 
and it is possible to explain their origins by the radiative breaking mechanism. 
This is because the soft mass squared of the singlet is naturally driven to be negative at the low-energy scale, 
leading to a non-zero vacuum expectation value. 
In this case, the required masses for the singlet and vector-like fields are generated without explicit mass terms in the superpotential.

\section*{Acknowledgments}
Grants-in-Aid for Scientific Research from the Ministry of Education, Culture, Sports, Science, and Technology (MEXT), Japan, No. 26104009 (T. T. Y.); Grant-in-Aid No. 26287039 (T. T. Y.) from the Japan Society for the Promotion of Science (JSPS); and by the World Premier International Research Center Initiative (WPI), MEXT, Japan (T. T. Y.).
The research leading to these results has received funding
from the European Research Council under the European Unions Seventh
Framework Programme (FP/2007-2013) / ERC Grant Agreement n. 279972
``NPFlavour'' (N.\,Y.).

\appendix

\section{Two-loop beta-functions} \label{ap:rges}

The two-loop RGEs of gauge couplings and gauginos with $N_5$ pairs of $\bf 5$ and $\bar{\bf 5}$ are 
\begin{eqnarray}
\beta_{g_i}^{(1)} &=& \frac{g_i^3}{16\pi^2} \left(b_i^{(1)} + N_5 \right), \nonumber \\
\beta_{g_i}^{(2)} &=& \frac{g_i^3}{(16\pi^2)^2}  \left(\sum_{j} b_{ij}^{(2)} g_j^2 - \sum_{k=t, b, \tau, L', D'} c_{ik}^{(2)} Y_k^2 \right), 
\end{eqnarray}
and
\begin{eqnarray}
\beta_{M_i}^{(1)} &=& \frac{2 g_i^2}{16\pi^2} \left(b_i^{(1)} + N_5 \right) M_i \, ,\nonumber \\
\beta_{M_i}^{(2)} &=& \frac{2g_i^2}{(16\pi^2)^2}\left(\sum_{j} b_{ij}^{(2)} g_j^2 (M_i + M_j) + \sum_{k=t, b, \tau, L', D'} c_{ik}^{(2)} Y_k^2 (-M_i + A_k )\right) ,
\end{eqnarray}
where $Y_{L'}$ and $Y_{D'}$ correspond to $\lambda_{L}$ and $\lambda_D$, respectively, $b_i^{(1)}=\{ 33/5, 1, -3\}$. The two-loop coefficients are
\begin{eqnarray}
b_{ij}^{(2)}&=& 
\left(
\begin{array}{ccc}
 \frac{199}{25} + \frac{7}{15} N_5 &   \frac{27}{5} + \frac{9}{5} N_5 & \frac{88}{5}  +  \frac{32}{15} N_5 \vspace{3pt}\\
 \frac{9}{5}+ \frac{3}{5} N_5 &   25+  7 N_5&   24  \vspace{3pt} \\  
  \frac{11}{5}+ \frac{4}{15} N_5& 9  &   14+ \frac{34}{3} N_5
\end{array}
\right),
\end{eqnarray}
\begin{eqnarray}
c_{ik}^{(2)}&=&
\left(
\begin{array}{ccccc}
 \frac{26}{5} &   \frac{14}{5}& \frac{18}{5}  & \frac{6}{5} N_5 & \frac{4}{5} N_5 \vspace{3pt} \\
  6&   6&   2 & 2 N_5 & 0\vspace{3pt} \\
  4&  4 &   0 & 0 & 2 N_5\vspace{3pt} 
\end{array}
\right).
\end{eqnarray}

Let us summarize beta-functions of the SUSY invariant parameters with $N_5=4$. 
We have used {\tt Susyno} package~\cite{susyno} for obtaining the following two-loop RGEs.
The notation is $\beta_i=\beta_i^{(1)}/(16\pi^2) +\beta_i^{(2)}/(16\pi^2)^2.$  
\begin{eqnarray}
\beta_{\lambda_L}^{(1)} &=& \lambda_L\Bigl(12\lambda_D^2 + 10\lambda_L^2 -\frac{3}{5}g_1^2- 3 g_2^2 \Bigr) ,\nonumber \\
\beta_{\lambda_L}^{(2)} &=& \lambda_L\Bigr[ 
-24 \lambda_D^4  -34 \lambda_L^4 -24 \lambda_L^2 \lambda_D^2 
+ \lambda_D^2 ( \frac{16}{5} g_1^2 + 64 g_3^2)
+ \lambda_L^2 ( \frac{24}{5} g_1^2 + 24 g_2^2) \nonumber \\
&+& \frac{327}{50} g_1^4 + \frac{9}{5} g_1^2 g_2^2 + \frac{39}{2} g_2^4
\Bigr], \nonumber \\
\beta_{\lambda_D} ^{(1)} &=& \lambda_D \Bigl(14\lambda_D^2 + 8\lambda_L^2 - \frac{4}{15} g_1^2 - \frac{16}{3} g_3^2 \Bigr), \nonumber \\
\beta_{\lambda_D} ^{(2)} &=& \lambda_D \Bigr[ 
-50 \lambda_D^4 -16 \lambda_D^2 \lambda_L^2 -16 \lambda_L^4
+ \frac{16}{5} \lambda_D^2 ( g_1^2 + 20 g_3^2)
+ \frac{24}{5} \lambda_L^2 ( g_1^2 + 5 g_2^2) \nonumber \\
&+& \frac{644}{225} g_1^4 + \frac{64}{45} g_1^2 g_3^2 + \frac{176}{9} g_3^4
\Bigr], \nonumber \\
\beta_{M_L}^{(1)} &=& M_L\Bigl( 2 \lambda_L^2 - \frac{3}{5}g_1^2 - 3 g_2^2 \Bigr), \nonumber \\
\beta_{M_L}^{(2)} &=& M_L \Bigr[
-18 \lambda_L^4 -24 \lambda_D^2 \lambda_L^2
+ \frac{327}{50} g_1^4 + \frac{9}{5} g_1^2 g_2^2 
+ \frac{39}{2} g_2^4
 \Bigr], \nonumber \\
\beta_{M_D}^{(1)} &=& M_D \Bigl( 2 \lambda_D^2 - \frac{4}{15}g_1^2 - \frac{16}{3} g_3^2 \Bigr), \nonumber \\
\beta_{M_D}^{(2)} &=& M_D \Bigr[
-26 \lambda_D^4 -16 \lambda_D^2 \lambda_L^2
+ \frac{644}{225} g_1^4 + \frac{64}{45} g_1^2 g_3^2 + \frac{176}{9} g_3^4
 \Bigr], \nonumber \\
\beta_{M_S}^{(1)} &=& M_S \Bigl(16 \lambda_L^2 + 24 \lambda_D^2 \Bigr), \nonumber \\
\beta_{M_S}^{(2)} &=& M_S\Bigl[ 
-48 \lambda_D^4 -32 \lambda_L^4 
+ \frac{32}{5} \lambda_D^2 (g_1^2 + 20 g_3^2)
+ \frac{48}{5} \lambda_L^2 (g_1^2 + 5 g_2^2)
\Bigr]. \nonumber \\
\end{eqnarray}

The beta-functions of the MSSM parameters are modified at the two-loop level:
\begin{eqnarray}
\delta \beta_{Y_u}^{(2)} &=& Y_u \Bigl( \frac{52}{15} g_1^4 + 12 g_2^4 + \frac{64}{3} g_3^4 \Bigr),\nonumber \\
\delta \beta_{Y_d}^{(2)} &=& Y_d \Bigl( \frac{28}{15} g_1^4 + 12 g_2^4 + \frac{64}{3} g_3^4\Bigr),\nonumber \\
\delta \beta_{Y_e}^{(2)} &=& Y_e \Bigl( \frac{36}{5} g_1^4 + 12 g_2^4 \Bigr), \nonumber \\
\delta \beta_{\mu}^{(2)} &=& \mu \Bigl( \frac{12}{5} g_1^4 + 12 g_2^4 \Bigr). \nonumber \\
\end{eqnarray}

Next, we show the two-loop RGEs of SUSY breaking mass parameters.
The beta-functions for new trilinear couplings are
\begin{eqnarray}
\beta_{A_{\lambda_L}}^{(1)} & =&   24 A_{\lambda_D} \lambda_D^2 + 20 A_{\lambda_L} \lambda_L^2 
+ \frac{6}{5} g_1^2 M_1 + 6 g_2^2 M_2 ,
         \nonumber \\
\beta_{A_{\lambda_L}}^{(2)} & =&          
-96 \lambda_D^4 A_{\lambda_D} - 136 \lambda_L^4 A_{\lambda_L}
-48 \lambda_L^2  \lambda_D^2 (A_{\lambda_D} + A_{\lambda_L})
\nonumber \\
&+& \lambda_D^2 \Bigl( \frac{32}{5} A_{\lambda_D} (g_1^2 + 20 g_3^2) 
- \frac{32}{5} g_1^2 M_1 - 128 g_3^2 M_3 \Bigr) \nonumber \\
&+& \frac{48}{5} \lambda_L^2 \Bigl(
A_{\lambda_L} (g_1^2 + 5 g_2^2) - g_1^2 M_1 - 5 g_2^2 M_2	
\Bigr) \nonumber \\
&-& \frac{654}{25} g_1^4 M_1 - \frac{18}{5} g_1^2 g_2^2 (M_1 + M_2)
-78 g_2^4 M_2
 ,\nonumber \\
\beta_{A_{\lambda_D}}^{(1)}   &=& 28 A_{\lambda_D} \lambda_D^2 + 16 A_{\lambda_L} \lambda_L^2 
+ \frac{8}{15} g_1^2 M_1 +\frac{32}{3}  g_3^2 M_3  , \nonumber \\
\beta_{A_{\lambda_D}}^{(2)} & =&   
-200 \lambda_D^4 A_{\lambda_D} - 64 \lambda_L^4 A_{\lambda_L}
-32 \lambda_L^2  \lambda_D^2 (A_{\lambda_D} + A_{\lambda_L})
\nonumber \\
&+& \lambda_D^2 \Bigl( \frac{32}{5} A_{\lambda_D} (g_1^2 + 20 g_3^2) 
- \frac{32}{5} g_1^2 M_1 - 128 g_3^2 M_3 \Bigr) \nonumber \\
&+& \frac{48}{5} \lambda_L^2 \Bigl(
A_{\lambda_L} (g_1^2 + 5 g_2^2) - g_1^2 M_1 - 5 g_2^2 M_2	
\Bigr) \nonumber \\
&-& \frac{2576}{225} g_1^4 M_1 - \frac{128}{45} g_1^2 g_3^2 (M_1 + M_3)
-\frac{704}{9} g_3^4 M_3
.
\end{eqnarray}

Beta-functions of trilinear and bilinear couplings in the MSSM have additional terms as
\begin{eqnarray}
\delta \beta_{A_u}^{(2)} & =&  - \frac{208}{15}g_1^4 M_1  - 48 g_2^4 M_2 - \frac{256}{3} g_3^4 M_3    ,      \nonumber \\
\delta \beta_{A_d}^{(2)}   &=& - \frac{112}{15}g_1^4 M_1  - 48 g_2^4 M_2 - \frac{256}{3}g_3^4 M_3      ,  \nonumber \\
\delta \beta_{A_e}^{(2)}    &=&   - \frac{144}{15}g_1^4 M_1  - 48 g_2^4 M_2   , \nonumber \\
\delta \beta_{B_H}^{(2)}    &=&  -\frac{48}{5} g_1^4 M_1 - 48 g_2^4 M_2 .
\end{eqnarray}

The beta-functions for the soft mass squared parameters of new particles are
\begin{eqnarray}
\beta_{m_S^2}^{(1)} &=& 8\left[  3 \lambda_D^2 ( A_{\lambda_D}^2 + m_{D'}^2 + m_{\bar D'}^2  + m_S^2 ) + 2 \lambda_L^2 (A_{\lambda_L}^2 + m_{L'}^2 + m_{\bar L'}^2  + m_S^2 )  \right] \, ,\nonumber \\
\beta_{m_S^2}^{(2)} &=&  -64 \lambda_L^4 ( 2 A_{\lambda_L}^2 + m_{L'}^2 + m_{\bar L'}^2 + m_S^2)
- 96 \lambda_D^4 (2 A_{\lambda_D}^2 + m_{D'}^2 + m_{\bar D'}^2 + m_S^2) \nonumber \\
&+& \frac{48}{5} \lambda_L^2 \Bigl[
(m_{L'}^2 + m_{\bar L'}^2 + m_S^2 + A_{\lambda_L}^2)  (g_1^2 + 5 g_2^2) \nonumber \\
&+& 2 (g_1^2 M_1^2 + 5 g_2^2 M_2^2) - 2 A_{\lambda_L} (g_1^2 M_1 + 5 g_2^2 M_2)
\Bigr] \nonumber \\
&+& \frac{32}{5} \lambda_D^2 \Bigl[
(m_{D'}^2 + m_{\bar D'}^2 + m_S^2 + A_{\lambda_D}^2)  (g_1^2 + 20  g_3^2) \nonumber \\
&+& 2 (g_1^2 M_1^2 + 20 g_3^2 M_3^2) - 2 A_{\lambda_D} (g_1^2 M_1 + 20 g_3^2 M_3)
\Bigr]\, ,
\nonumber \\
\beta_{m_{\bar D'}^2}^{(1)} &=&   2 \lambda_D^2 (A_{\lambda_D}^2 + m_{D'}^2 + m_{\bar D'}^2 + m_S^2) 
+\frac{2}{5} g_1^2 S  \, , \nonumber \\
\beta_{m_{\bar D'}^2}^{(2)} &=&  
-52 \lambda_D^4 (2 A_{\lambda_D}^2 + m_{D'}^2 + m_{\bar D'}^2 + m_S^2) \nonumber \\
&-& 16 \lambda_D^2 \Bigl[
\lambda_L^2 ( (A_{\lambda_D} +  A_{\lambda_L})^2 
+ m_{D'}^2 + m_{\bar D'}^2 + m_{L'}^2 + m_{\bar L'}^2 + 2 m_S^2)  \nonumber \\
&+&\frac{g_1^2}{5} m_{\bar D'}^2 - \frac{g_1^2}{5} m_{D'}^2
\Bigr]
+ \lambda_L^2 g_1^2 \left(\frac{16}{5} m_{L'}^2 -\frac{16}{5} m_{\bar L'}^2 \right) \nonumber \\
&-&\frac{6}{5} g_1^2 g_2^2 \Bigl[m_{H_d}^2- m_{H_u}^2 + \sum_i (m_{L_i}^2 - m_{Q_i}^2) 
+ 4 (m_{L'}^2 - m_{\bar L'}^2) \Bigr] \nonumber \\
&+&\frac{32}{45} g_1^2 g_3^2 
\Bigl[
\sum_i (3 m_{\bar D_i}^2 + 3 m_{Q_i}^2 - 6 m_{\bar U_i}^2) + 12 m_{\bar D'}^2 - 12 m_{D'}^2  
+ 4 M_1^2  + 4 M_1 M_3 + 4 M_3^2
\Bigr] \nonumber \\
&+& \frac{2}{75} g_1^4 \Bigl[
 \sum_i (8 m_{\bar D_i}^2 + 48 m_{\bar E_i}^2 - 3 m_{L_i}^2  + 3 m_{Q_i}^2
-16 m_{\bar U_i}^2) - 3 m_{H_d}^2 + 15 m_{H_u}^2 \nonumber \\
&+& 32 m_{\bar D'}^2  -12 m_{L'}^2 + 60 m_{\bar L'}^2 + 644 M_1^2
\Bigr]
\nonumber \\
&+& \frac{4}{5} g_1^2 S_1  + \frac{16}{3}g_3^4 S_3 \, , \nonumber \\
\beta_{m_{D'}^2}^{(1)} &=& 2 \lambda_D^2 (A_{\lambda_D}^2 + m_{D'}^2 + m_{\bar D'}^2 + m_S^2)
-\frac{2}{5} g_1^2 S \, ,
 \nonumber \\
\beta_{m_{D'}^2}^{(2)} &=&  
-52 \lambda_D^4 (2 A_{\lambda_D}^2 + m_{D'}^2 + m_{\bar D'}^2 + m_S^2) \nonumber \\
&-& 16 \lambda_D^2 \Bigl[
\lambda_L^2 ( (A_{\lambda_D} + A_{\lambda_L})^2 
+ m_{D'}^2 + m_{\bar D'}^2 +  m_{L'}^2 + m_{\bar L'}^2 + 2 m_S^2)  \nonumber \\
&-&\frac{g_1^2}{5} m_{\bar D'}^2 + \frac{g_1^2}{5} m_{D'}^2
\Bigr] - \lambda_L^2 g_1^2 \left(\frac{16}{5} m_{L'}^2 -\frac{16}{5} m_{\bar L'}^2 \right) \nonumber \\
&+&\frac{6}{5} g_1^2 g_2^2 \Bigl[m_{H_d}^2- m_{H_u}^2 + \sum_i (m_{L_i}^2 - m_{Q_i}^2) 
+ 4 (m_{L'}^2 - m_{\bar L'}^2) \Bigr] \nonumber \\
&-&\frac{32}{45} g_1^2 g_3^2 
\Bigl[
\sum_i (3 m_{\bar D_i}^2 + 3 m_{Q_i}^2 - 6 m_{\bar U_i}^2) + 12 m_{\bar D'}^2 - 12 m_{D'}^2  
- 4 M_1^2  - 4 M_1 M_3 - 4 M_3^2
\Bigr] \nonumber \\
&+& \frac{2}{75} g_1^4 \Bigl[
 \sum_i ( -24 m_{\bar E_i}^2  + 15 m_{L_i}^2  +  m_{Q_i}^2
 + 48 m_{\bar U_i}^2) +15 m_{H_d}^2 -3 m_{H_u}^2 \nonumber \\
&+& 32 m_{D'}^2  + 60 m_{L'}^2 -12 m_{\bar L'}^2 + 644 M_1^2
\Bigr] \nonumber \\
&-& \frac{4}{5} g_1^2 S_1 + \frac{16}{3}g_3^4 S_3  \, ,
\nonumber \\
\beta_{m_{L'}^2}^{(1)} &=&  2 \lambda_L^2 (A_{\lambda_L}^2 + m_{L'}^2 + m_{\bar L'}^2 + m_S^2)
-\frac{3}{5} g_1^2 S  \, ,
\nonumber \\
\beta_{m_{L'}^2}^{(2)} &=& 
-36 \lambda_L^4 ( 2 A_{\lambda_L}^2 + m_{L'}^2 + m_{\bar L'}^2 + m_S^2) \nonumber \\
&+& \lambda_D^2 \Bigl[
-24 \lambda_L^2 ( (A_{\lambda_D} + A_{\lambda_L})^2 + 
m_{\bar D'}^2 + m_{D'}^2 + m_{L'}^2 + m_{\bar L'}^2 + 2 m_S^2) \nonumber \\
&+& \frac{24}{5} g_1^2 (m_{\bar D'}^2 - m_{D'}^2)
\Bigr] 
-\frac{24}{5} \lambda_L^2 g_1^2 ( m_{L'}^2 - m_{\bar L'}^2 ) \nonumber \\
&+& \frac{9}{5} g_1^2 g_2^2 
\Bigl[
m_{H_d}^2 - m_{H_u}^2 + \sum_i (m_{L_i}^2 - m_{Q_i}^2) + 4 (m_{L'}^2 - m_{\bar L'}^2)
+ 2 (M_1 + M_2)^2
\Bigr] \nonumber \\
&-&\frac{16}{5} g_1^2 g_3^2 \Bigl[
\sum_i (m_{\bar D_i}^2 + m_{Q_i}^2 - 2 m_{\bar U_i}^2 )
+ 4 (m_{\bar D'}^2 - m_{D'}^2)
\Bigr] \nonumber \\
&+& \frac{1}{25} g_1^4 \Bigl[
\sum_i (2 m_{\bar D_i}^2  + 2 m_{Q_i}^2  + 56 m_{\bar U_i}^2 - 18 m_{\bar E_i}^2 + 18 m_{L_i}^2 )
+18 m_{H_d}^2 \nonumber \\
&+& 8 m_{\bar D'}^2 + 40 m_{D'}^2 + 72 m_{L'}^2 + 981 M_1^2
\Bigr] \nonumber \\
&-& \frac{6}{5} g_1^2  S_1 + 3 g_2^4 S_2 \, , \nonumber \\
\beta_{m_{\bar L'}^2}^{(1)} &=& 2 \lambda_L^2 (A_{\lambda_L}^2 + m_{L'}^2 + m_{\bar L'}^2 + m_S^2)
+\frac{3}{5} g_1^2 S \, ,
\nonumber \\
\beta_{m_{\bar L'}^2}^{(2)} &=& 
-36 \lambda_L^4 ( 2 A_{\lambda_L}^2 + m_{L'}^2 + m_{\bar L'}^2 + m_S^2) \nonumber \\
&+& \lambda_D^2 \Bigl[
-24 \lambda_L^2 ( (A_{\lambda_D} + A_{\lambda_L})^2 + 
m_{\bar D'}^2 + m_{D'}^2 + m_{L'}^2 + m_{\bar L'}^2 + 2 m_S^2) \nonumber \\
&-& \frac{24}{5} g_1^2 (m_{\bar D'}^2 - m_{D'}^2)
\Bigr] + \frac{24}{5} \lambda_L^2 (m_{L'}^2 - m_{\bar L'}^2)\nonumber \\
&-& \frac{9}{5} g_1^2 g_2^2 
\Bigl[
m_{H_d}^2 - m_{H_u}^2 + \sum_i (m_{L_i}^2 - m_{Q_i}^2) + 4 (m_{L'}^2 - m_{\bar L'}^2)
- 2 (M_1 + M_2)^2
\Bigr] \nonumber \\
&+&\frac{16}{5} g_1^2 g_3^2 \Bigl[
\sum_i (m_{\bar D_i}^2 + m_{Q_i}^2 - 2 m_{\bar U_i}^2 )
+ 4 (m_{\bar D'}^2 - m_{D'}^2)
\Bigr] \nonumber \\
&+& \frac{1}{25} g_1^4 \Bigl[
\sum_i (10 m_{\bar D_i}^2  + 4 m_{Q_i}^2  -8 m_{\bar U_i}^2 + 54 m_{\bar E_i}^2 )
+18 m_{H_u}^2 \nonumber \\
&+& 40 m_{\bar D'}^2 + 8 m_{D'}^2 + 72 m_{\bar L'}^2 + 981 M_1^2
\Bigr] \nonumber \\
&+& \frac{6}{5} g_1^2 S_1 + 3 g_2^4 S_2 ,\nonumber \\
\end{eqnarray}
where 
\begin{eqnarray}
S_3 &=& \sum_i (2 m_{Q_i}^2 + m_{\bar D_i}^2 + m_{\bar U_i}^2 ) + 4 (m_{D'}^2 + m_{\bar D'}^2)  + 16 M_3^2
\, ,\nonumber \\
S_2 &=& \sum_i (3 m_{Q_i}^2 + m_{L_i}^2 ) 
+ m_{H_u}^2 + m_{H_d}^2 + 4 (m_{L'}^2 + m_{\bar L'}^2)  + 35 M_2^2
\, ,
\nonumber \\
S_1 &=& (3 Y_b^2 + Y_{\tau}^2) m_{H_d}^2 + Y_{\tau}^2 (- 2 m_{\bar E_3}^2 + m_{L_3}^2)
-Y_b^2 (2 m_{\bar D_3}^2 + m_{Q_3}^2)  \nonumber \\
&-& Y_t^2 (3 m_{H_u}^2 + m_{Q_3}^2 -4 m_{\bar U_3}^2)
\, ,
\nonumber \\
S&=& m_{H_u}^2 - m_{H_d}^2 +
{\rm Tr}\left[ m_{Q}^2 - m_{L}^2 - 2 m_{\bar U}^2 + m_{\bar D}^2 + m_{\bar E}^2 \right] +
\delta S \, ,
\end{eqnarray}
with $\delta S = 4( - m_{L'}^2 + m_{\bar L'}^2 + m_{\bar D'}^2 - m_{D'}^2 )$.

The additional contributions to one-loop beta-functions for soft mass squared parameters of MSSM fields are 
\begin{eqnarray}
\delta \beta_{m_{Q}^2 }^{(1)}&=&\frac{1}{5} \delta S \, , \ \ 
\delta \beta_{m_{\bar U}^2 }^{(1)}=-\frac{4}{5} \delta S\, , \, \ \ 
\delta \beta_{m_{\bar D}^2 }^{(1)}=\frac{2}{5} \delta S \, ,\nonumber \\
\delta \beta_{m_{L}^2 }^{(1)}&=&-\frac{3}{5} \delta S \, , \, \ \ 
\delta \beta_{m_{\bar E}^2 }^{(1)}= \frac{6}{5} \delta S \, ,\nonumber \\
\delta \beta_{m_{H_u}^2 }^{(1)}&=&\frac{3}{5} \delta S \, \, ,\ \ 
\delta \beta_{m_{H_d}^2 }^{(1)}=-\frac{3}{5} \delta S\, . \nonumber \\
\end{eqnarray}
The two-loop beta-functions of the soft mass squared parameters are modified as 
\begin{eqnarray}
\delta \beta_{m_{Q}^2 }^{(2)}&=& 
- \frac{8}{5} \lambda_D^2 g_1^2 ( m_{\bar D'}^2 - m_{D'}^2 )
+ \frac{8}{5} \lambda_L^2 g_1^2 ( m_{L'}^2 - m_{\bar L'}^2 ) \nonumber \\
&+& \frac{64}{15} g_1^2 g_3^2 (m_{\bar D'}^2 - m_{D'}^2) 
- \frac{12}{5} g_1^2 g_2^2 (m_{L'}^2 - m_{\bar L'}^2) \nonumber \\
&+& \frac{8}{75} g_1^4 \Bigl(
3 m_{\bar D'}^2 - m_{D'}^2 - 3 m_{L'}^2 + 6 m_{\bar L'}^2 + 15 M_1^2
\Bigr) \nonumber \\
&+& 3 g_2^4 \delta S_2
+ \frac{16}{3} g_3^4 \delta S_3 \, ,
\nonumber \\
\delta \beta_{m_{\bar U}^2 }^{(2)}&=& 
\frac{32}{5} \lambda_D^2 g_1^2 (m_{\bar D'}^2 - m_{D'}^2) 
-\frac{32}{5} \lambda_L^2 g_1^2 (m_{L'}^2 - m_{\bar L'}^2)  \nonumber \\
&-& \frac{256}{15} g_1^2 g_3^2 \Bigl( 
m_{\bar D'}^2 - m_{D'}^2
\Bigr) 
+ \frac{48}{5} g_1^2 g_2^2 \Bigl( 
m_{L'}^2 - m_{\bar L'}^2
\Bigr) \nonumber \\
&+& \frac{16}{75} g_1^4 \Bigl(
4 m_{\bar D'}^2 + 12 m_{D'}^2 + 21 m_{L'}^2 + 3 m_{\bar L'}^2 + 120 M_1^2
\Bigr)
\nonumber \\
&+& \frac{16}{3} g_3^4 \delta S_3 \, ,
\nonumber \\
\delta \beta_{m_{\bar D}^2 }^{(2)}&=& 
-\frac{16}{5} \lambda_D^2 g_1^2 (m_{\bar D'}^2 - m_{D'}^2) 
+\frac{16}{5} \lambda_L^2 g_1^2 (m_{L'}^2 - m_{\bar L'}^2)  \nonumber \\
&+& \frac{128}{15} g_1^2 g_3^2 (m_{\bar D'}^2 -m_{D'}^2)
-  \frac{24}{5} g_1^2 g_2^2 (m_{L'}^2 -m_{\bar L'}^2) \nonumber \\
&+& \frac{8}{75} g_1^4 \Bigl( 8 m_{\bar D'}^2 - 3 m_{L'}^2 + 15 m_{\bar L'}^2  + 60 M_1^2\Bigr) 
\nonumber \\
&+& \frac{16}{3} g_3^4 \delta S_3 \, ,
\nonumber \\
\delta \beta_{m_{L}^2 }^{(2)}&=& 
\frac{24}{5} \lambda_D^2 g_1^2 (m_{\bar D'}^2 - m_{D'}^2) 
-\frac{24}{5} \lambda_L^2 g_1^2 (m_{L'}^2 - m_{\bar L'}^2)  \nonumber \\
&-& \frac{64}{5} g_1^2 g_3^2 (m_{\bar D'}^2 -m_{D'}^2)
+  \frac{36}{5} g_1^2 g_2^2 (m_{L'}^2 -m_{\bar L'}^2) \nonumber \\
&+& \frac{8}{25} g_1^4 \Bigl(m_{\bar D'}^2 + 5 m_{D'}^2 + 9 m_{L'}^2  + 45 M_1^2\Bigr) \nonumber \\
&+& 3 g_2^4 \delta S_2  \, ,
\nonumber \\
\delta \beta_{m_{\bar E}^2 }^{(2)} &=&
-\frac{48}{5} \lambda_D^2 g_1^2 (m_{\bar D'}^2 - m_{D'}^2) 
+\frac{48}{5} \lambda_L^2 g_1^2 (m_{L'}^2 - m_{\bar L'}^2)  \nonumber \\
&+& \frac{128}{5} g_1^2 g_3^2 (m_{\bar D'}^2 -m_{D'}^2)
-  \frac{72}{5} g_1^2 g_2^2 (m_{L'}^2 -m_{\bar L'}^2) \nonumber \\
&+& \frac{8}{25} g_1^4 \Bigl( 16 m_{\bar D'}^2 + 8 m_{D'}^2 +  9 m_{L'}^2 +  27 m_{\bar L'}^2  + 180 M_1^2\Bigr) 
\, ,
\nonumber \\
\delta \beta_{m_{H_u}^2 }^{(2)}&=& 
-\frac{24}{5} \lambda_D^2 g_1^2 (m_{\bar D'}^2 - m_{D'}^2) 
+\frac{24}{5} \lambda_L^2 g_1^2 (m_{L'}^2 - m_{\bar L'}^2)  \nonumber \\
&+& \frac{64}{5} g_1^2 g_3^2 (m_{\bar D'}^2 -m_{D'}^2)
-  \frac{36}{5} g_1^2 g_2^2 (m_{L'}^2 -m_{\bar L'}^2) \nonumber \\
&+& \frac{8}{25} g_1^4 \Bigl( 5 m_{\bar D'}^2 +  m_{D'}^2 + 9 m_{\bar L'}^2  + 45 M_1^2\Bigr) \nonumber \\
&+& 3 g_2^4 \delta S_2 \, ,
\nonumber \\
\delta \beta_{m_{H_d}^2 }^{(2)} &=& 
\frac{24}{5} \lambda_D^2 g_1^2 (m_{\bar D'}^2 - m_{D'}^2) 
-\frac{24}{5} \lambda_L^2 g_1^2 (m_{L'}^2 - m_{\bar L'}^2)  \nonumber \\
&-& \frac{64}{5} g_1^2 g_3^2 (m_{\bar D'}^2 -m_{D'}^2)
+  \frac{36}{5} g_1^2 g_2^2 (m_{L'}^2 -m_{\bar L'}^2) \nonumber \\
&+& \frac{8}{25} g_1^4 \Bigl(m_{\bar D'}^2 + 5 m_{D'}^2 + 9 m_{L'}^2  + 45 M_1^2\Bigr) \nonumber \\
&+& 3 g_2^4 \delta S_2 \, ,
\end{eqnarray}
where
\begin{eqnarray}
\delta S_3 = 4 m_{\bar D'}^2 + 4 m_{D'}^2 + 24 M_3^2 \, , \nonumber \\
\delta S_2 = 4 m_{L'}^2 + 4 m_{\bar L'}^2 + 24 M_2^2 \, .
\end{eqnarray}

Beta-functions of new $B$-terms are 
\begin{eqnarray}
\beta_{B_{S}}^{(1)} & =&  16 (3 A_{\lambda_D} \lambda_D^2 + 2 A_{\lambda_L} \lambda_L^2) ,         \nonumber \\
\beta_{B_{S}}^{(2)} &=& \Bigl[ 
-192 \lambda_D^4 A_{\lambda_D} 
- 128 \lambda_L^4 A_{\lambda_L}  \nonumber \\
&+& \frac{64}{5}\lambda_D^2 (g_1^2 + 20 g_3^2) A_{\lambda_D} 
- \frac{64}{5}\lambda_D^2 (g_1^2 M_1 + 20 g_3^2 M_3)    \nonumber \\
&+& \frac{96}{5}\lambda_L^2 (g_1^2 + 5 g_2^2)  A_{\lambda_L}
- \frac{96}{5}\lambda_L^2 (g_1^2 M_1 + 5 g_2^2 M_2 ) 
\Bigr], \nonumber \\
\beta_{M_L}^{(1)} &=& \Bigl( 4 \lambda_L^2 A_{\lambda_L} + \frac{6}{5}g_1^2 M_1 + 6 g_2^2 M_2 \Bigr), \nonumber \\
\beta_{M_L}^{(2)} &=& \Bigl[ 
-72\lambda_L^4 A_{\lambda_L} 
-48\lambda_L^2\lambda_D^2(A_{\lambda_L} + A_{\lambda_D}) \nonumber \\
&-& \frac{654}{25} g_1^4 M_1 - \frac{18}{5} g_1^2 g_2^2(M_1+M_2) - 78 g_2^4 M_2 
\Bigr], \nonumber \\
\beta_{M_D}^{(1)} &=&  \Bigl( 4 \lambda_D^2A_{\lambda_D}+ \frac{8}{15}g_1^2 M_1 + \frac{32}{3} g_3^2 M_3 \Bigr),\nonumber \\
\beta_{M_D}^{(2)} &=& \Bigl[ 
-104 \lambda_D^4 A_{\lambda_D} 
-32 \lambda_L^2\lambda_D^2(A_{\lambda_L} + A_{\lambda_D}) \nonumber \\
&-& \frac{2576}{225}g_1^4 M_1 - \frac{128}{45}g_1^2 g_3^2(M_1+M_3) - \frac{704}{9} g_3^4 M_3 
\Bigr]. \nonumber \\
\end{eqnarray}

\end{document}